# An Efficient Algorithm for Cooperative Spectrum Sensing in Cognitive Radio Networks

Vinod Sharma and ArunKumar Jayaprakasam
Dept of Electrical Communication Engineering
Indian Institute of Science, Bangalore, 560012
Email: {vinod@ece.iisc.ernet.in, jarun@ti.com}

*Abstract*— We consider the problem of Spectrum Sensing in Cognitive Radio Systems. We have developed a distributed algorithm that the Secondary users can run to sense the channel cooperatively. It is based on sequential detection algorithms which optimally use the past observations. We use the algorithm on secondary users with energy detectors although it can be used with matched filter and other spectrum sensing algorithms also. The algorithm provides very low detection delays and also consumes little energy. Furthermore it causes low interference to the primary users. We compare this algorithm to several recently proposed algorithms and show that it detects changes in spectrum faster than these algorithms and uses significantly less energy.

## I. INTRODUCTION

THE Electromagnetic Radio Spectrum, a natural resource, is currently licensed by regulatory bodies for various applications. Presently there is a severe shortage of the spectrum for new applications and systems. However various studies ([2], [3]) have concluded that at any time and place, very little of the licensed spectrum is actually utilized. The unutilized part of the spectrum results in '*Spectrum holes*' or '*White Spaces*'. Therefore, recently it has been proposed to allow utilization of the unused spectrum at a time to other users who do not hold the license. This will be possible by the Cognitive Radio technology being developed now ([4], [7], [13]) Cognitive Radios are defined as radio systems which perform Radio Environment analysis, identify the spectrum holes and then operate in those holes.

In cognitive radio terminology *Primary user* refers to a user who is allocated the rights to use the spectrum. *Secondary user* refers to the users who try to use the frequency bands allocated to primary user when the primary user is not using it.

*Spectrum Sensing*, an essential component of the Cognitive Radio technology involves, 1) identifying spectrum holes and 2) when an identified spectrum hole is being used by the secondary users, to quickly detect the onset of primary transmission. This needs to be done such that guaranteed interference levels to the primary are maintained and there is efficient use of spectrum by the secondary. This involves detecting reliably, quickly and robustly, possibly weak, primary user signals.

In practice Spectrum Sensing becomes a challenging task because the channel from the primary transmitter to the secondary user can be bad because of *Shadowing* and time varying multipath *Fading*. As a result, detecting the primary user based on the observation of a single secondary user may not be enough, especially under low SNR conditions. Hence to alleviate this problem, Cooperative Spectrum Sensing ([3], [10], [19]) is envisaged, whereby the spatial diversity inherent in radio environment is leveraged by allowing multiple secondary users to cooperate. This reduces the average time to detect the primary user. This in turn lowers the interference to the primary user (when it switches ON), while increasing the spectrum usage of the secondary's (when the primary switches OFF). Since multiple geographically distributed secondary nodes are involved, efficient decentralized cooperation protocols are required to effectively utilize the power and bandwidth [14]. In the next section we describe briefly different Spectrum Sensing algorithms available in the literature.

### A. Literature Survey

In [7] various aspects of Cognitive Radio technology are covered. For spectrum sensing, primarily three Signal Processing Techniques ([2], [3]) are proposed in literature: Matched filter [15], Energy Detection [15] and cyclo-stationary feature detection [20]. Matched filtering is optimal. However it requires detailed knowledge of primary signaling. But when no knowledge about the primary user signal is assumed, an Energy Detector is Optimal [15]. Hence most of the papers are based on energy detection. Also a new standard of IEEE on cognitive radio (IEEE 802.22) will provide for spectrum sensing via energy detection.

The limitations of Local Spectrum Sensing and the benefits of Cooperative Spectrum Sensing are illustrated in [19]. In [19] the performance of fusion rules – AND, OR and MAJORITY is compared under different SNR profiles of the secondary nodes. It is shown that when one node has much higher SNR compared to other nodes then cooperative spectrum sensing performs worse than the single node spectrum sensing. Thus in [8], the fusion center takes a weighted sum of the local decisions to arrive at the global decision, the weights being proportional to the secondary channels' SNRs.

In [14] the fusion center takes optimal weighted average of the

This research is partially supported by a grant from Boeing Corporation.

received data from different users and arrives at the final decision. In [6] Spectrum sensing is considered in a cooperative communication context. Two secondary users cooperate using the Amplify and Forward (AF) protocol. In [9] a confidence level based voting scheme is proposed to decrease the amount of energy utilized in performing spectrum sensing, while keeping the performance loss negligible. In [16] a dual threshold scheme is proposed to reduce the bandwidth requirements to perform spectrum sensing.

The above studies detect the spectrum hole from a single snapshot of observations from one or more nodes. In [1] we developed an efficient decentralized algorithm to detect changes in the distribution of a stochastic sequence. Our algorithm is based on Sequential Detection techniques, in particular on CUSUM [12]. It is known that sequential detection techniques perform better than the optimal rules based on a single observation. These techniques gainfully utilize the past observations along with the current. Recently these algorithms have been used in the decentralized setup also (see [1], [11], [17] and the references therein). These can be implemented online, are iterative in nature and require minimal computations at each step. In this paper we use the algorithm in [1] for cooperative spectrum sensing. We show that this algorithm performs better than other recent algorithms mentioned above.

This paper is organized as follows. Section II describes the model and the DualCUSUM algorithm developed in [1]. In Section III we compare this algorithm with those in [6], [9], [15] and [19] and show that it performs better than those algorithms. Section IV concludes the paper.

## II. MODEL AND ALGORITHM

### A. Model

Consider a Cognitive Radio System with one primary transmitter and $L$ Secondary users. The observations made on the channel by the $L$ secondaries are used by a fusion center to make a decision whether the channel is free or not.

The secondary nodes have to detect the change in the status of the channel in two situations. First, when the primary has been using the channel and it stops transmission. The secondary nodes have to cooperatively detect this change as soon as possible so that they can make maximum use of the available channel. The second situation is when the channel has been free and the secondaries are using the channel. They need to sense the channel to see if the primary starts transmission. It is difficult to detect this when the secondaries are transmitting, especially when the energy detection method is used. Thus, in between their transmissions, secondary's stop intermittently for a few slots and sense the channel to see if primary has started transmission. If yes the secondary's need to stop using that channel. To minimize interference to the primary, this also needs to be done quickly.

We present our algorithm in the setup of the second scenario but it can be effectively used for the first scenario as well. At the end we will briefly comment on it. Our algorithm is based on the energy detection method as this has been the most studied model. However our algorithm can be modified for use in the matched filter / cyclostationary feature detection methods also.

Let hypothesis $H_0$ refers to the scenario when the primary user is not transmitting and $H_1$ when the primary is transmitting. At the $k^{th}$ time instant the signal received by the $l^{th}$ user is, $X_{k,l} = N_{k,l}, l = 1,2,...,L$, under hypothesis $H_0$ and $X_{k,l} = h_l S_k + N_{k,l}, l = 1,2,...,L$, under hypothesis $H_1$ where $h_l$ is the channel gain of the $l^{th}$ user, $S_k$ is the primary user signal and $N_{k,l}$ is observation noise at the $l^{th}$ user. We assume a slow fading model whereby the channel gain is random, but remains constant throughout the spectrum sensing session. We also assume that the primary uses a constant modulation scheme, and that $\{N_{k,l}, k = 1,2,...\}$ are i.i.d.

The primary user is assumed to start transmission at a random time $T$. The aim is to detect this change at the fusion center as soon as possible at a time τ (say) using the messages transmitted from the $L$ Sensors. Each of the nodes $l$ uses its observation $X_{k,l}$ to generate a signal $Y_{k,l}$ and transmits to the Fusion Center. The data received at the fusion center is corrupted by i.i.d noise $Z_k$. The fusion center uses $Y_{k,1}, Y_{k,2},...,Y_{k,L}$ to decide if the primary is transmitting at time $k$ or not. Thus each of the slots reserved for sensing the channel by the secondaries is further split into two parts. In the first part the secondaries sense the channel and in the second part they transmit $Y_{k,l}$. Also to transmit $Y_{k,1}, Y_{k,2},...,Y_{k,L}$ they will need to use some MAC protocol. For energy detection, TDMA is the most used protocol.

Let $P_{FA} = P(\tau < T)$ be the probability that the fusion center decides that the primary has started transmitting while it has not. We state the goal of the problem as:

min $EDD \triangleq E[(\tau - T)^+]$

Subject to: $P_{FA} = P(\tau < T) \leq \alpha$   (1)

where $\alpha$ is an appropriately chosen constant. This seems a reasonable formulation for our application and is commonly used in sequential detection of change literature. However in spectrum sensing literature the formulation used in many studies is ([6], [14], [18]): maximize the probability $p_d$ of detecting $H_1$ in a slot given that primary is transmitting in that slot, with a constraint on $p_{fa}$, the probability of detecting $H_1$ in a slot when the primary is not transmitting in that slot, namely $p_{fa} \leq \alpha'$.

If $T$, the time the primary starts transmitting is geometrically distributed, i.e., $P[T=n]=(1-\rho)^{n-1}\rho, 0<\rho<1$ then the two formulations can be shown to be equivalent if $p_d$ and $p_{fa}$ are not changing from slot to slot (which is usually assumed in literature). This is because then,

$$p_{fa} = \frac{P_{FA}\rho}{(1-P_{FA})(1-\rho)} \quad (2)$$

and $E[(\tau-T)^+] = 1/p_d$. Thus minimizing $EDD$ is equivalent to maximizing $p_d$. Also $p_{fa}$ is a strictly monotonically increasing function of $P_{FA}$. Thus at the optimal point $P_{FA} = \alpha$ and $p_{fa} = \alpha'$ where $\alpha$ and $\alpha'$ are related by obtaining $\alpha'$ from (2) by substituting $\alpha$ for $P_{FA}$.

In the following we will use DualCUSUM with $T$ geometric and hence the DualCUSUM which provides a good solution of (1) also provides a good solution of (2).

Actually the assumption of constant $p_d$ and $p_{fa}$ is satisfied in the other studies on spectrum sensing mentioned above. But CUSUM and DualCUSUM increase $p_d$ with time. This makes DualCUSUM a better algorithm.

The DualCUSUM algorithm in [1] has several advantages over the above setup. It uses all past observations at each secondary user optimally to make the current decision. Also it reduces the average detection delay $EDD$ and hence also increases the spectrum usage by the secondaries. All secondary nodes transmit their observations $Y_{k,1}, Y_{k,2} ... Y_{k,L}$ in the same slot, i.e., one slot is used in transmission instead of $L$ slots. We assume that there is synchronization of these symbols at the fusion node so that it senses $\sum Y_{k,l} + Z_k$, where $Z_k$ is the receiver noise at the fusion node. In addition unlike in other algorithms, in DualCUSUM a secondary transmits $Y_{k,l}$ only when it decides locally via CUSUM if the change has occurred and not otherwise. This saves transmission energy at the secondary and also reduces interference to the primary in case it has already started transmission. Finally even the fusion node uses CUSUM to optimally utilize all the past data from all the secondary's to make a decision. As a result of these several desirable features we will see that DualCUSUM substantially outperforms other existing algorithms in spectrum sensing. In the following we describe the DualCUSUM algorithm.

*B. DualCUSUM Algorithm*

1) Each of the Secondary users runs Parametric CUSUM algorithm:
   $W_{k,l} = \max(0, W_{k-1,l} + \xi_{k,l}), \ W_{0,l} = 0$
   $where, \ \xi_{k,l} = \log[f_{1,l}(X_{k,l})/f_0(X_{k,l})].$

   Here $f_{1,l}$ is the density of $X_{k,l}$ under $H_1$ and $f_0$ is the density of $X_{k,l}$ under $H_0$.

2) Secondary user $l$ transmits at time $k$, only if $W_{k,l} > \gamma$. If the threshold is exceeded it transmits a value $b$ i.e. $Y_{k,l} = b 1_{\{W_{k,l} > \gamma\}}$. The parameters $b$ and $\gamma$ are chosen appropriately. This step allows to save energy and cause less interference to others.

3) At Fusion Center we assume Physical Layer Fusion:
   $Y_k = \sum Y_{k,l} + Z_k$
   where $Z_k$ is i.i.d noise at the fusion node.

4) Change Detection at Fusion Center via CUSUM:
   $F_k = \max\{0, F_{k-1} + \log\frac{g_I(Y_k)}{g_0(Y_k)}\}$
   where $g_0$ is the density of $Z_k$ and $g_I$ is the density of $Z_k + bI$, $I$ being a design parameter.

5) The Fusion Center declares a change at time $\tau(\beta,\gamma,b,I)$ when $F_k$ crosses a threshold $\beta$:
   $\tau(\beta,\gamma,b,I) = \inf\{k: F_k > \beta\}$

The CUSUM algorithm used at the secondary node and at the fusion center has negative drift before change and positive drift after change. Thus as soon as the change occurs (i.e., primary starts transmitting) the CUSUM process starts increasing and quickly crosses the threshold to declare change.

In the above algorithm we have assumed that the channel from the secondary users to the fusion center has no fading although that can also be taken care of. The performance parameters $P_{FA}$ and $EDD$ critically depend on the parameters $(\beta,\gamma,b,I)$. In [1] $P_{FA}$ and $EDD$ are analytically computed for each $(\beta,\gamma,b,I)$ and an iterative algorithm is designed to optimize these parameters. Finally the same DualCUSUM algorithm works if we want to detect the time when the primary stops the transmission, possibly with different parameters.

As mentioned above the DualCUSUM algorithm was designed for sensor networks where energy saving is of prime importance. It was achieved in DualCUSUM by making sensor nodes transmit to the fusion node only when they detect change in local CUSUM. In the present context also this step provides significant energy saving. Although most of the studies mentioned above do not worry about energy saving, some do ([9]). This step also reduces the interference to the primary (in case it is transmitting at that time).

DualCUSUM, even while saving energy significantly, performs substantially better than other algorithms in $EDD$. If energy is not of much concern, we can improve over DualCUSUM. For example we can let each secondary transmit its observations in each slot. Also physical layer fusion may or may not be exploited. If we let each secondary transmit in the same slot, we save in number of transmitted slots (thus reducing $EDD$) but we lose information. One option is to let each node transmit in a separate slot. Then one can use global CUSUM, i.e., as if the original observation of each secondary is available at the same place. This of course (except the effect of $L$ transmission slots) is globally optimal.

In the next section we will use only DualCUSUM except in III.D where we also use globalCUSUM to compare against another centralized Algorithm.

### III. COMPARISON WITH OTHER ALGORITHMS

In this section we compare the performance of DualCUSUM with some of the other cooperative spectrum sensing algorithms available in the literature. For all algorithms we meet a target $P_{FA}$ and obtain the $EDD$.

#### A. Gaussian Case

In this section we assume that the pre-change distribution $f_0$ and post-change distribution $f_1$ are Gaussian. This is valid when the decision statistic is the received power expressed in dB and log-normal distribution is used for modeling the shadowing ([18]). We compare the performance of DualCUSUM with some simple slot-based fusion rules. Here each secondary node compares the received power with a threshold and accordingly decides a 1 ($H_1$) or 0 ($H_0$). Finally each node transmits the decision to the fusion center. The fusion center declares change at time τ according to one of the following three fusion rules:

- **OR**: Change is declared if any of the secondary decides 1
- **AND**: Change is declared if all the secondaries decides 1
- **MAJORITY**: Change is declared if majority of the secondaries decides 1.

Parameters used for comparison are based on [18] and are as follows. There are six secondary nodes. Noise is assumed to be Gaussian with mean 0 and variance 1 in each of the secondary nodes and at the fusion center. The post change means in the nodes are 0.5, 0.9, 1.1, 0.3, 1.5, 0.75. $T$ is assumed to be Geometric (ρ=0.01). For the DualCUSUM algorithm $b$ is fixed at 3.1623. $I$ is chosen to be 5 as that gave the best results for us. $EDD$ for each algorithm is in Table 1.

| $EDD$ | $P_{FA}$=0.1 | $P_{FA}$=0.027 | $P_{FA}$=0.01 |
|---|---|---|---|
| OR | 24.9260 | 73.4785 | 154.2 |
| AND | 14.6451 | 34.9357 | 63.4647 |
| MAJORITY | 9.1071 | 22.8804 | 43.6 |
| DualCUSUM | 3.6553 | 5.0933 | 5.9856 |

**Table 1: EDD for DualCUSUM vs. Simple Fusion (Gaussian)**

One sees that DualCUSUM performs much better than other algorithms. The difference increases as $P_{FA}$ decreases. For practical systems one expects low $P_{FA}$.

#### B. Energy Detection

Next we compare the performance of DualCUSUM with the above algorithms without making the Gaussian approximation. We also compare with Confidence Level (CL) algorithm developed in [9] with MAJORITY fusion. We only assume that the observation noises are i.i.d Gaussian with zero mean and variance 1 in each of the secondary nodes and the fusion center. In this scheme each secondary node calculates the energy of the last $N$ samples i.e. the detection statistic is $E_{K,l} = \sum_{k=1}^{N} X_{k,l}^2$. Then the computed energy is compared with a threshold. If the threshold is exceeded decide $H_1$ otherwise $H_0$. For a single slot, to meet a given target probability of false alarm, the threshold can be computed from results available in [5]. For DualCUSUM, now $f_0$ is central chi-square distribution of order $N$ and $f_1$ is non-central chi-square distribution of order $N$ and parameter $N|h_l|^2$. Here although for simplicity we assume $S_k$ is 1, it is sufficient that $\sum_{k=1}^{N}|S_k|^2 = N$, i.e., instead of assuming constant modulus if its average over N samples is approximately constant.

Parameters used for comparisons are as in [14]. There are six secondary users each with primary to secondary channel gain being -3.7, -5.2, -3.4, -5.4, -9.5, -3.8 dB respectively. $N$ = 20 samples are used to compute the energy. $T$ is assumed to be Geometric (ρ=0.05), $b$ =3.1623 and $I$ =5.

| $EDD$ | $P_{FA}$=0.1 | $P_{FA}$=0.027 | $P_{FA}$=0.01 |
|---|---|---|---|
| OR | 5.2674 | 13.904 | 25.8454 |
| AND | 4.5480 | 9.3234 | 15.304 |
| MAJORITY | 2.2942 | 5.0638 | 8.2840 |
| MAJORITY+CL | 2.344 | 5.16 | 8.54 |
| DualCUSUM | 1.7766 | 2.5966 | 3.25 |

**Table 2: DualCUSUM vs. Simple Energy Detectors**

In Table 2 EDD is expressed in units of $N$ = 20 slots. Among the three fusion rules MAJORITY performs the best, which is in accordance with [19]. As target $P_{FA}$ decreases, gain due to DualCUSUM becomes substantial over the other algorithms.

#### C. Cooperative Spectrum Sensing

In this section we compare the performance of DualCUSUM, with the spectrum sensing algorithm in [6]. There are two secondary nodes which cooperate using amplify and forward (AF) protocol to get an agility gain, i.e., reduction in EDD. The channel gains are 0dB and 5dB respectively. The overall noise variance is assumed to be unity. For DualCUSUM $b$ is chosen as 3.1623. Performance comparison is provided in Table 3.

| EDD | $P_{FA}$=0.1 | $P_{FA}$=0.027 | $P_{FA}$=0.01 |
|---|---|---|---|
| Cooperative | 6.22 | 12.8 | 19.4 |
| DualCUSUM | 3.25 | 4.71 | 5.5443 |

**Table 3: DualCUSUM vs. Cooperative Spectrum Sensing**

*D. Optimal Linear Cooperation*

In the 'optimal linear cooperation' algorithm [14], each of the secondary nodes transmits the exact energy received to the fusion center. At the fusion center, the energy from all the nodes is combined linearly using an optimal weight factor. We provide the comparison in Table 4. The data for [14] is obtained by simulating the MDC (Modified Deflection Coefficient) based algorithm in [14] whose performance is close to the optimal linear cooperation. The parameters used are the same as in Section III.B.

| EDD | $P_{FA}$=0.1 | $P_{FA}$=0.027 | $P_{FA}$=0.01 |
|---|---|---|---|
| MDC | 1.0063 | 2.25 | 3.5683 |
| DualCUSUM | 1.7766 | 2.5966 | 3.25 |
| GlobalCUSUM | 0.8034 | 1.3359 | 1.7 |

**Table 4: DualCUSUM vs. Linear Cooperation**

As can be seen performance of DualCUSUM is inferior to Optimal Linear Cooperation at high $P_{FA}$. But as $P_{FA}$ decreases performance of DualCUSUM becomes better. However, Linear Cooperation is a *centralized* detection algorithm using the global knowledge of the received SNR at each secondary node, noise variance at each node and noise variance at the fusion node. Also it consumes lot more energy than DualCUSUM (see Table 5). Thus for a fair comparison we also implemented GlobalCUSUM which uses the same energy and information as in [14]. *EDD* for the GlobalCUSUM is also provided in Table 4. We see that it performs better than the DualCUSUM and Linear Cooperation.

*E. Energy Saving Comparison*

In this section we obtain the total energy needed for transmission in DualCUSUM algorithm and compare against the Optimal Linear Cooperation and MAJORITY+CL [9]. In Table 5 ETR denotes the average number of transmissions per node (which is directly related to transmission energy and the interference caused to the primary) till the fusion center declares change. The parameters used are the same as in Section III.B.

| ETR | $P_{FA}$=0.1 | $P_{FA}$=0.027 | $P_{FA}$=0.01 |
|---|---|---|---|
| Majority+CL | 18.8333 | 24.167 | 28.38 |
| Linear cooperation | 18.75 | 21.6226 | 23.4762 |
| Dual-CUSUM | 2.38 | 2.1526 | 1.9833 |

**Table 5: Energy Saving of DualCUSUM**

## IV. CONCLUSIONS AND FUTURE WORK

We used the DualCUSUM algorithm developed in [1] and compared the performance with other cooperative spectrum sensing algorithms. It performs better than all decentralized algorithms in mean delay to detect as well as the energy consumed. DualCUSUM uses knowledge of the density of observations. The only problem in knowing this is in knowing the fading value which is random. More recently we have modified DualCUSUM to take care of this which will be reported elsewhere. The performance of the modified algorithm is close to DualCUSUM.